\documentclass{kluwer}
\usepackage{epsfig}
\newcommand{\R}{${\cal R}$} 
\newcommand{\etal}{{\it et al.\ }}
\newcommand{\lta}{\stackrel{<}{\scriptstyle\sim}}
\newcommand{\gta}{\stackrel{>}{\scriptstyle\sim}}
\begin{document}
\begin{article}

\begin{opening}
\title{On the Nature and Redshift Evolution of DLA Galaxies}
\runningtitle{DLA Galaxies}
\author{Fritze - v. Alvensleben, \surname{U.}\email{ufritze@uni-sw.gwdg.de}} 

\author{Lindner, \surname{U.}}
\author{M\"oller, \surname{C.S.}}
\author{Fricke, \surname{K.J.}}
\institute{Universit{\"a}ts -- Sternwarte G{\"o}ttingen}

\begin{abstract}
We extend our spiral galaxy models that successfully describe nearby 
template spectra as well as the redshift evolution of CFRS and HDF 
spirals to include -- in a chemically consistent way -- the redshift evolution 
of a series of individual elements. 
Comparison with observed DLA abundances shows that DLAs might well 
be the progenitors of present-day spiral types Sa through Sd. 
Our models bridge the gap between high redshift DLA and nearby spiral 
HII region abundances. The slow redshift evolution of DLA abundances 
is a natural consequence of the long SF timescales for disks, the scatter 
at any redshift reflects the range of SF timescales from early to 
late spiral types. We claim that while at high redshift all spiral
progenitor types seem to give rise to DLA absorption, towards low
redshifts, the early type spirals seem to drop out of DLA samples 
due to low gas and/or high metal and dust content. 
Model implications for the spectrophotometric properties of the DLA
galaxy population are discussed in the context of campaigns for the
optical identifications of DLA galaxies both at low and high redshift. 

\end{abstract}
\end{opening}


\section{Introduction}
The basis for the work presented here is our set of chemically
consistent spiral galaxy models with a range of Star Formation
Histories ({\bf SFHs}) specifying the spectral types $Sa,~Sb,~Sc,~Sd$.
Our models are very simple 1-zone models without any dynamics
meant to describe global quantities of average galaxies of the
respective types. Our approach is to use the simplest models possible
with the smallest number of parameters in order to see how far we can
get and what kind of sophistications are required by a comparison with
observed galaxy properties.

Our unified spectrophotometric and chemical evolutionary synthesis
models allow to have a large number of observational constraints --
spectrophotometric properties including gaseous emission and stellar
absorption features as well as ISM abundances -- to restrict the small
number of model parameters, basically the IMF and the SFH. 
While in the local Universe a 1--1 - correspondance between spectral
types and morphological types is a long-standing matter of fact it is
also clear that this correspondance has to break down at some yet
unknown stage when going back towards the earliest phases of galaxy
evolution and formation.

The choice of SFHs for the various spectral types is determined by a
number of observational constraints. These include type-averaged
luminosities, colours from UV through NIR for nearby galaxy samples,
emission and absorption line properties, template spectra, and HII
region abundances that our model galaxies have to match after a Hubble
time as well as a comparison with observed galaxy luminosities and
colours over a large redshift range (cf. M\"oller \etal 1999). 

Here, we present the chemical evolution aspect of our spectroscopically
successful spiral galaxy models and compare to abundances for
a series of elements observed in Damped Ly$\alpha$ ({\bf
DLA}) Absorbers over the redshift range from ${\rm z \sim 0.4}$
through ${\rm z \gta 4.4}$, which corresponds to lookback times of
more than 90 \% of the age of the Universe. Originally, DLA absorption
was thought to arise in intervening (proto-)galactic disks along the
lines of sight to distant QSOs (Wolfe 1988).
This view is supported by arguments
based on mass estimes from column densities and absorber sizes at
${\rm z \sim 2 - 3}$ as well as by kinematic features consistent with
rotation at velocities of order 200 ${\rm km~s^{-1}}$ (Prochaska \&
Wolfe 1997a, b, 1998). More recently, on the basis of [$\alpha$/Fe] and
[N/O] abundance ratios and
their large scatter, an origin of DLA absorption in dwarf or low
surface brightness galaxies is
also discussed  (Matteucci \etal 1997, Vladilo 1998, Jimenez \etal 1999).
Our aim is to see, if and in how far very simple spiral galaxy models
are compatible with the observed DLA abundance evolution over the
large redshift range accessible. In this context it does not matter if
DLA galaxies at the highest redshifts are not yet fully assembled
massive disks but rather consist of galactic building blocks as
suggested by Haehnelt \etal (1998). Our SFHs in this case are meant to
describe SF in all the fragments bound to end up in one disk by z $=
0$. 

In a second step, we then present the spectrophotometric properties of
those galaxy models that succesfully describe the DLA galaxy
population and discuss them in the context of the large campaigns
designed to optically identify DLA galaxies both at low and at high
redshift. 

\section{A Chemically Consistent Chemical Evolution Model}
As opposed to star clusters which basically form their stars ``all at the
same time'', i.e. within $\sim 10^5$ yr, any stellar system with a SFH
more extended than this is expected to feature finite distributions
both in age and metallicity. Our spectrophotometric and chemical
evolutionary synthesis models are {\bf chemically consistent} in the
sense that we account for the increasing initial metallicity of
successive generations of stars. We keep track of the ISM abundance
at birth of each star and use various sets of input physics for
metallicities in the range ${\rm -2.5 \leq [Fe/H] \leq +0.3}$. In
particular, we use stellar evolutionary tracks and lifetimes from the
Padova group (Bressan \etal 1993, Fagotto \etal 1994 a, b, c 
for 0.6 ${\rm \leq  m_{\ast} \leq 120~M_{\odot}}$, and from 
Chabrier \& Baraffe 1997 for ${\rm m _{\ast} \leq 0.5~M_{\odot}}$,
stellar yields and remnant masses from  v. d. Hoek \& Groenewegen
(1997) for ${\rm m_{\ast} \leq 8~M_{\odot}}$ and from  Woosley \&
Weaver (1995) for stars ${\rm 12 \leq m_{\ast} \leq 40~M_{\odot}}$,
and model atmosphere spectra and colour calibrations from
Lejeune \etal (1998). SNIa contributions are included as
described by Matteucci \& Greggio (1986) and Matteucci \& Tornamb\`e
(1987) with SNIa yields from Nomoto \etal (1997).

\noindent
We use a Salpeter IMF and SFHs for the various spiral types as
follows:
\begin{center} $Sa~ ...~ Sc$ \hspace{3.cm} $~~{\rm \Psi(t) \sim
\frac{G}{M}(t)}$ \\
$~~~Sd$ \hspace{3.4cm}  ${\rm \Psi(t) \sim const.}$\end{center}
(G: gas mass, M: total mass).
Characteristic timescales for SF ${\rm t_{\ast}}$ as defined by
$\int_0^{t_{\ast}} \Psi \cdot dt = 0.63 \cdot G (t=0)$ are 

\begin{center}
 \begin{tabular}{rrc} 
   $~~t_{\ast}~ \sim~ $ &2 \ Gyr &$ ~~Sa~~$ \\
   $~~t_{\ast}~ \sim~ $ &3 \ Gyr &$ ~~Sb~~$ \\
   $~~t_{\ast}~ \sim~ $ &10 \ Gyr &$ ~~Sc~~$ \\
   $~~t_{\ast}~ \sim~ $ &16 \ Gyr &$ ~~Sd~~$ \\ 
 \end{tabular}
\end{center}

Our models have a strong analytic power in the sense that they allow
to trace back the luminosity contributions to any wavelength band as
well as the enrichment contributions to any chemical element of
different stellar masses, spectral types, luminosity classes,
metallicity subpopulations, nucleosynthetic origins (PNe, SNI, SNII,
single stars, binaries, ...) as a function of time or redshift.

While for the spectro-cosmological evolution cosmological and
evolutionary corrections as well as the effect of attenuation have to
be considered, the chemo-cosmological evolution simply results from a
1-1-transformation of galaxy age into redshift (we use a standard
cosmology ${\rm (H_o,~\Omega_o,~\Lambda_o) =
(75,~1,~0)}$ and assume galaxy formation at ${\rm z_{form}
= 5}$).   

The above SFHs were chosen as to give agreement after a Hubble time of
evolution with average colours for the galaxy types $Sa,~Sb,~Sc,$ $Sd$
from the RC3 and template spectra from Kennicutt (1992).  
In most nearby spirals HII region abundances show negative gradients
with galactocentric radius and even at a given radius within a galaxy,
there may be considerable scatter among abundances of individual HII
regions. Geometrical considerations by Phillipps \& Edmunds (1996) and
Edmunds \& Phillipps (1997) show
that for an arbitrary sightline featuring DLA
absorption 
the most probable galactocentric distance to pass through an
intervening galaxy disk  is around ${\rm 1~R_e}$.
At the same time, ${\rm 1~R_e}$
seems to be a reasonable radius for HII region abundances to compare to our
global 1-zone models. The agreement of average HII region abundances
for various galaxy types at ${\rm 1~R_e}$ as measured by Zaritsky
\etal (1994), Oey \& Kennicutt (1993), van Zee \etal (1998), and Ferguson
(1998) with model ISM abundances after a Hubble time confirms our
choice of SFHs. 


\section{Abundance Evolution of Spiral Galaxy Models and DLA Observations}
High resolution
spectra are required to fully resolve the velocity structure and
derive precise heavy element abundances in DLA absorption systems. In
recent years, a large number of precise abundances have been
determined for elements C, N, O, Al, Si, S, Cr, Mn, Fe, Ni, Zn, ... in
a large number of DLAs over the redshift range ${\rm z \sim 0.4}$
through
${\rm z \gta 4.4}$ (Boiss\'e \etal 1998, Lu \etal 1993, 1996,
Pettini \etal 1994, 1999, Prochaska \& Wolfe 1997, de la
Varga \& Reimers 1999). For the comparison with the redshift
evolution of our models we
have carefully referred all published DLA abundances to one
homogeneous set of oscillator strengths and solar reference values
(see Lindner \etal 1999 for details).

\begin{figure}
\centerline{\epsfig{file=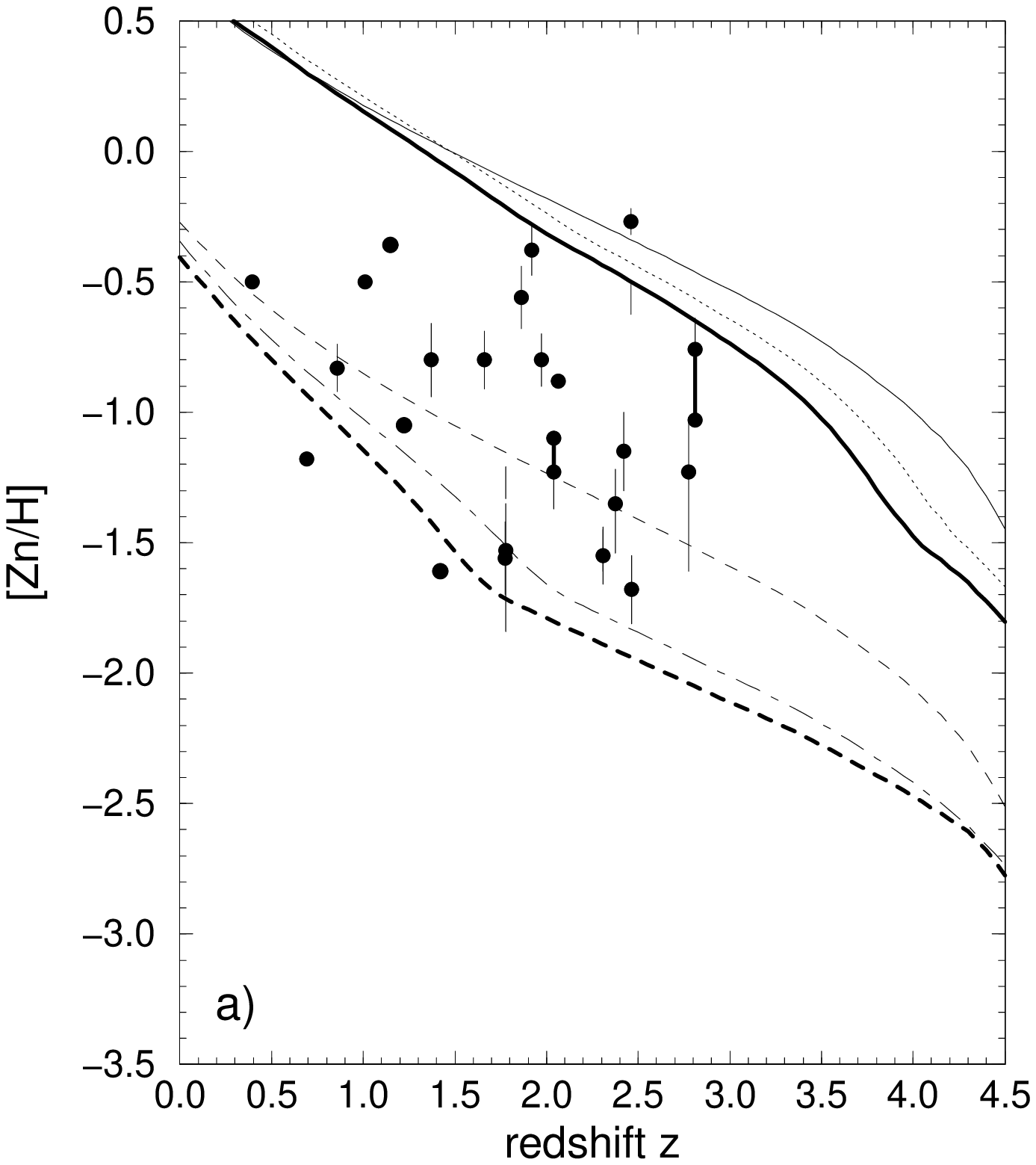,width=15pc}\epsfig{file=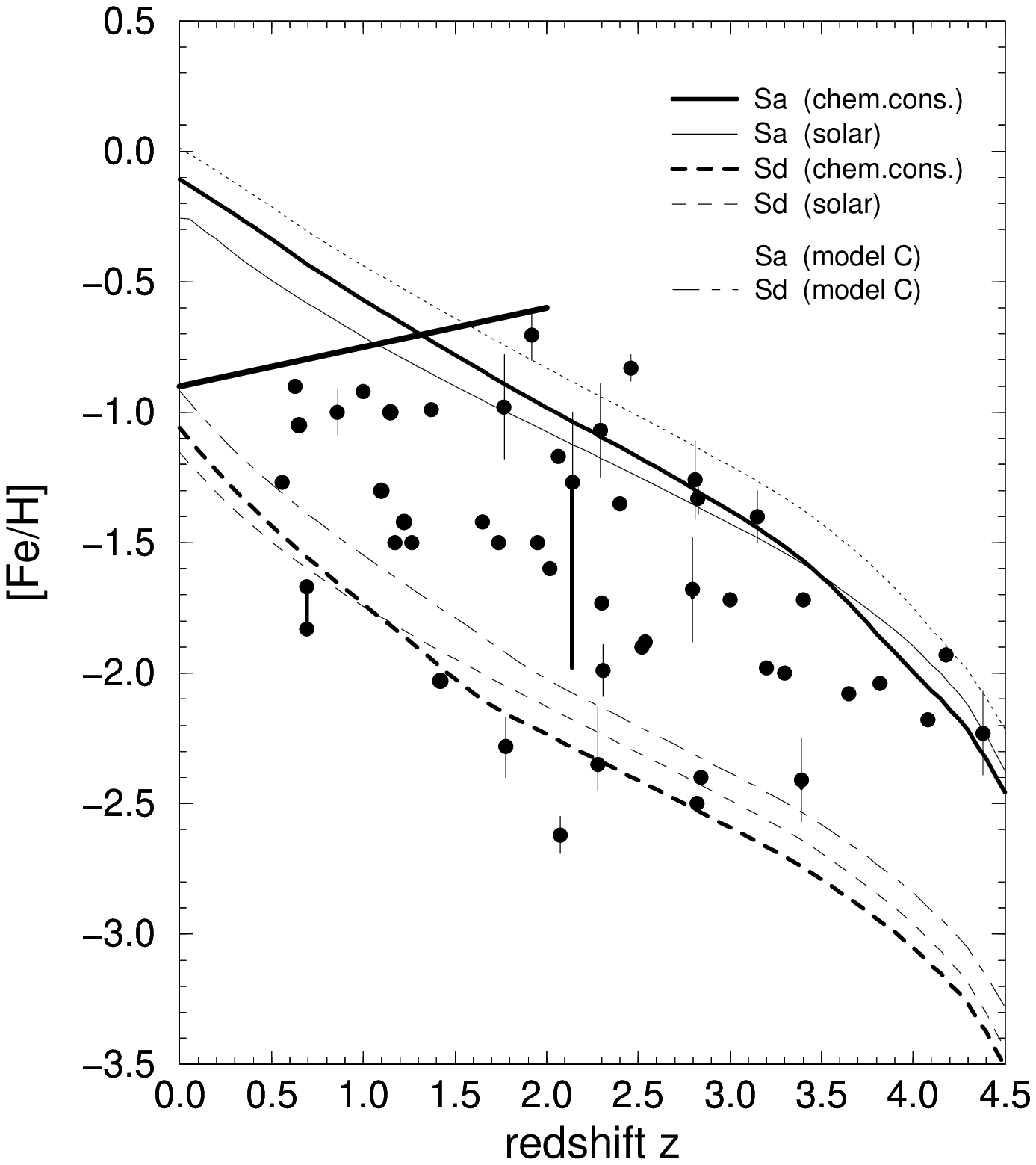,
width=15pc}}
\caption{Redshift evolution of [Zn/H] (1a) and [Fe/H] (1b) for models
 Sa and Sd compared to observed DLA abundances.}
\end{figure}

Fig. 1 shows the redshift evolution of our ``slowest evolution'' $Sd$
and our ``fastest evolution'' $Sa$ models in comparison with the
available [Zn/H] (1a) and [Fe/H] (1b) abundances in DLAs. $Sb$ and
$Sc$ models are omitted to avoid confusion, their abundance evolution
runs in between those of the $Sa$ and $Sd$ models. Lindner \etal
(1999) present a more extensive comparison including all those 8 elements
for which abundances are measured in a reasonable number of DLAs.

As compared to models using solar metallicity input physics only (thin
lines), the
chemically consistent models (heavy lines) show significant
differences ranging up to 1.0 dex for some elements. Changes in the
explosion energy (Woosley \& Weaver's model C) are considerably less
important. It should be noted that no scaling or additional parameters are 
introduced. Using SFHs, IMF, and yields, as desribed above, results in
absolute model abundances.

The conclusions we draw from this comparison (including all elements)
are the following:
\begin{itemize}
\item $Sa$ -- $Sd$ models bracket the redshift evolution of DLA abundances
from ${\rm z \geq 4.4}$ to ${\rm z \sim 0.4}$,
\item models bridge the gap from high-z DLAs  to 
nearby spiral HII region abundances, 
\item the weak redshift evolution of DLA abundances is a natural result of the 
long SF timescales for disks galaxies,
\item the range of SF timescales 
$t_{\ast}$ for near-by spirals from $Sa$ through $Sd$ fully explains the abundance 
scatter among DLAs at fixed redshift. 
\end{itemize}
This means that from the point of view of abundance evolution over
more than 90\% of the age of the Universe normal spiral galaxies (or
their progenitors) are perfectly consistent with DLA
observations. This does not exclude the possibility that some
starbursting dwarfs or LSBs may also give rise to DLA absorption.
Moreover, as indicated by the heavy line in Fig. 1b, the early type
spirals that well seem to be present in high-z DLA samples seem to
drop out of low-z DLA samples.
At the same time as delimiting the region in abundance vs. redshift
space where DLA absorbers seem to occur, the heavy line in
Fig. 1b. corresponds to a gas-to-total mass ratio of 0.5 in our galaxy
models. Indeed, low-z DLAs are observed to generally have low N(HI) 
(e.g. Lanzetta \etal 1997).

Our prediction from the comparison of models with data is that while
at high redshift (${\rm z \gta 2}$) the DLA absorbing galaxy
population may well comprise early as well as late type (proto-)spirals, the
low redshift (${\rm z \lta 1.5}$) DLA absorption systems seem to
have their origin in late-type gas-rich and metal-poor spirals. 
A bias against high metallicity and/or dust content in low-z DLA
absorbing galaxies has been suspected e.g. by Steidel \etal (1997). 
In our view, the
drop of the global gas content that clearly goes together with increasing
metallicity in nearby spirals may add yet another reason: at low gas
content the probability is significantly reduced that an arbitrary 
sightline through the galaxy goes through a region with high enough 
HI column density for
DLA absorption. 


\section{The Spectrophotometric Aspect of DLA Galaxies}
If confirmed by a larger sample of low-z DLAs, this prediction has
important implications for the possibility of optical
identifications. Locally, early type spirals are brighter by $\sim
1.5$ mag in $B$ on average than late-type spirals despite the fact
that any spiral type shows a range in luminosity with considerable
overlap between different types (e.g. Sandage \etal 1985).

We can now use the spectro-cosmological aspect of our evolutionary models as
e.g. presented by M\"oller \etal ({\sl this volume}) to predict the
range of apparent magnitudes and colours expected for DLA absorbing
galaxies at various redshifts.
We find the intriguing result that in all three bands B, \R~, and K the
intrinsically brighter early-type spirals show
almost the same apparent luminosities at ${\rm z \sim 2 - 3}$ as the 
intrinsically fainter average
late-type $Sd$ galaxies at ${\rm z \sim 0.5}$.

\begin{center}
  \begin{tabular}{|l|cc||cc|} \hline
  & \multicolumn{2}{|c|}{\bf Sa} & \multicolumn{2}{|c|}{\bf Sd} \\ \hline 
	& ~~z$~\sim$ 0.5~~ & ~~z$~\sim$ 2 -- 3 ~~& ~~z$~\sim$ 0.5~~ & ~~z$~\sim$ 2 -- 3 ~~\\ \hline 
	~~B~~ & 22.5 & 24 -- 25 & 25.5 & 29 -- 30.5  \\ 
      ~~\R~ ~~& 21 & 24 & 24.5 & 29 \\
	~~K~~ & 18.5 & 21.4 -- 22 & 22 & 26 -- 27 \\ \hline
  \end{tabular}
\end{center}

We thus would {\bf not} expect low-z DLA galaxies to be more easily
identified than at least the brighter ones among the high-z DLAs. 
On the basis of these luminosities we understand the non-detection of
DLA galaxies at z $\sim 3$ down to \R~ $\sim 25.5$ (Steidel \etal
1998) and the small number of DLA candidates (2/10) at ${\rm 1.5 \leq z
\leq 2.5}$ to K $\sim 21$ by Aragon - Salamanca \etal (1996). 
The galaxies identified by Steidel \etal (1994, 1995) as candidates
for DLA absorbers at ${\rm z_{abs} = 0.6922,~ 0.3950,~ 0.7568,~ and ~
0.8596}$ indeed have luminosities ${\rm -19.5 \lta M_B \lta -19.0}$
typical of late-type spirals. However, for their sample of 7 DLA
identifications in the range ${\rm 0.4 \lta z \lta 1}$ Le Brun \etal (1997)
find a variety of morphologies: spirals as well as compact and LSB 
objects. At high redshift, Djorgovski \etal (1996) and Djorgovski (1998) report
identifications of a bright disk galaxy (${\rm M_B \sim -20.4}$) with
SFR ${\rm \gta 8 M_{\odot}yr^{-1}}$ in agreement with our models at
${\rm z \sim 3.15}$, and of an ${\rm M_B \sim -19.5}$ galaxy with low
SFR at ${\rm z \sim 4.1}$. The two early-type DLA galaxies ($S0$)
put foreward by Lanzetta \etal (1997)
at ${\rm z = 0.16377}$ and by Miller \etal (1999) (NGC 4203 at ${\rm z
= 0}$) both show counterrotating gas disks. HI mapping shows NGC 4203
to be abnormally gas-rich (van Driel \etal 1988). These two cases
raise the issue that recent accretion or merging may provide
favourable conditions for DLA
absorption.


\section{Conclusions and Outlook}
We use very simple spiral galaxy models with standard IMF and SFHs
chosen as to agree with chemical and spectrophotmetric properties of
nearby galaxies as well as with the observed redshift evolution of 
luminosities and colours. When combined with a chemically consistent
description of their chemical evolution and a standard cosmology, 
we find good agreement with the observed redshift evolution of 
DLA abundances over more than 90 \% of the Hubble time. We claim that
at ${\rm z \gta 1.5}$ all spiral types can give rise to DLA absorption
while towards lower redshifts only the gas-rich metal-poor late spiral
types do so. The spectrophotometric properties given by our models for
average spiral types are consistent with the few optical
identifications of DLA galaxies both at low and high-z in large
observing programs.

Clearly, these simple models are a first approach only, a more
realistic treatment including infall and dynamical evolution has to
follow. 


\begin{acknowledgements}

UFvA and CSM gratefully acknowledge partial financial support to
attend the Conference. 
We warmly thank the organisers for a particularly fruitful and
inspiring Conference.
\end{acknowledgements}


\end{article}


\begin{thebibliography}{}


\bibitem{}{}{} Aragon -- Salamanca, A., Ellis, R. S., O'Brien, K. S., 1996, MN 281, 945

\bibitem{}{}{} Boiss\'e, P., Le Brun, V., Bergeron, J., Deharveng, J.-M., 1998, A\&A 333, 841

\bibitem{}{}{} Bressan, A., Fagotto, F., Bertelli, G., Chiosi, C., 1993, A\&AS 100, 647

\bibitem{} Le Brun, V., Bergeron, J., Boiss\'e, P., Deharveng, J. M.,
1997, A\&A 321, 733
\bibitem{}{}{} Chabrier, G., Baraffe, I., 1997, A\&A 327, 1039

\bibitem{} Djorgovski, S. G., 1997, in {\sl Structure and Evolution of
the IGM from QSO Absorption Lines} eds. P. Petitjean, S. Charlot,
Editions Fronti\`eres, p. 303

\bibitem{} Djorgovski, S. G., Pahre, M. A., Bechtold, J., Elston, R.,
1996, Nat 382, 234 

\bibitem{} van Driel, W., van Woerden, H., Gallagher, J.S., Schwarz,
U.J., 1988, A\&A 191, 201

\bibitem{} Edmunds, M. G., Phillipps, S., 1997, MN 292, 733

\bibitem{}{}{} Fagotto, F., Bressan, A., Bertelli, G., Chiosi, C., 1994 A\&AS 104, 365

\bibitem{}{}{} Fagotto, F., Bressan, A., Bertelli, G., Chiosi, C., 1994 A\&AS 105, 29 $+$ 39

\bibitem{}{}{} Ferguson, A. M. N., Gallagher, J. S., Wyse, R. F. G., 1998, AJ 116, 673

\bibitem{} Haehnelt, M. G., Steinmetz, M., Rauch, M., 1998, ApJ 495, 647

\bibitem{}{}{} v. d. Hoek, L. B., Groenewegen, M. A. T., 1997, A\&AS 123, 305

\bibitem{} Jimenez, R., Bowen, D. V., Matteucci, F., 1999, ApJ 514, L83

\bibitem{}{}{} Kennicutt, R. C., 1992, ApJS 79, 255
 

\bibitem{}{}{} Lanzetta, K. M., Wolfe, A. M., Altan, H., \etal, 1997,
AJ, 114, 1337

\bibitem{}{}{} Lejeune, T.,  Cuisinier, F., Buser, R., 1998, A\&AS 125, 229

\bibitem{}{}{} Lindner, U., Fritze -- v. Alvensleben, U., Fricke, K. J., 1999, A\&A 341, 709

\bibitem{}{}{} Lu, L., Wolfe, A. M., Turnshek, D. A., Lanzetta, K. M., 1993, ApJS 84, 1

\bibitem{}{}{} Lu, L., Sargent, W. L. W., Barlow, T. A., Churchill, C. W., Vogt, S. S., 1996, ApJS 107, 475

\bibitem{} Matteuci, F., Greggio, L., 1986, A\&A 154, 279

\bibitem{} Matteuci, F., Tornamb\`e, A., 1987, A\&A 185, 51

\bibitem{} Matteucci, F., Molaro, P., Vladilo, G., 1997, A\&A 321,
45


\bibitem{} Miller, E. D., Knezek, P.M., Bregman, J. N., 1999, ApJ 510, L95

\bibitem{}{}{} M\"oller, C. S., Fritze -- v. Alvensleben, U., Fricke, K. J., 1997, STScI Symp. Ser. 11

\bibitem{}{}{} M\"oller, C. S., Fritze -- v. Alvensleben, U., Fricke,
K. J., 1999, in {\sl The Birth of Galaxies}, {\sl in press}

\bibitem{}{}{} Nomoto, K., Iwamoto, K., Nakasto, N., \etal, 1997, Nucl. Phys. A, A621

\bibitem{}{}{} Oey, M. S., Kennicutt, R. C., 1993, ApJ 411, 137

\bibitem{}{}{} Pettini, M., Smith, L. J., Hunstead, R. W., King, D. L., 1994, ApJ 426, 79

\bibitem{}{}{} Pettini, M., Ellison, S. L., Steidel, C. C., Bowen,
D. V., 1999, ApJ, 510, 576

\bibitem{}  Phillipps, S., Edmunds, M. G., 1996, MN 281, 362

\bibitem{} Prochaska, J. X., Wolfe, A. M., 1997a, ApJ 474, 140

\bibitem{} Prochaska, J. X., Wolfe, A. M., 1997b, ApJ 487, 73

\bibitem{} Prochaska, J. X., Wolfe, A. M., 1998, ApJ 507, 113

\bibitem{}{}{} Sandage, A., Binggeli, B., Tammann, G. A., 1985, AJ 90, 395 $+$ 1759

\bibitem{}{}{} Steidel, C. C., Pettini, M., Dickinson, M., Persson,
S. E., 1994, AJ 108, 2046
\bibitem{}{}{} Steidel, C. C., Bowen, D. V., Blades, J. C.,
Dickinson, M., 1995, ApJ 440, L45
\bibitem{}{}{} Steidel, C. C., Dickinson, M., Meyer, D. M.,
Adelberger, K. L., Sembach, K. R., 1997, ApJ 480, 568
\bibitem{}{}{} Steidel, C. C., Adelberger, K. L., Dickinson, M.,
Giavalisco, M., Pettini, M., Kellogg, M., 1998, ApJ 492, 428
\bibitem{} de la Varga, A., Reimers, D., 1999, in {\sl Chemical
Evolution from Zero to High Redshift}, {\sl in press}

\bibitem{} Vladilo, G., 1998, ApJ 493, 583

\bibitem{} Wolfe, A. M., 1988, in {\sl QSO Absorption Lines: Probing the
Universe}, eds. C. Blades, D. Turnshek, C. Norman, Cambridge
University Press, p. 297

\bibitem{}{}{} Woosley, S. E., Weaver, T. A., 1995, ApJS 101, 181

\bibitem{}{}{} Zaritsky, D., Kennicutt, R. C., Huchra, J. P., 1994, ApJ 420, 87

\bibitem{} van Zee, L., Salzer, J. J., Haynes, M. P., O'Donoghue, A. A.,
Balonek, T. J., 1998, AJ 116, 2805

\end{thebibliography}
\end{document}